\documentclass[a4paper, amsfonts, amssymb, amsmath, reprint, showkeys, nofootinbib, twoside]{revtex4-1}
\usepackage[english]{babel}
\usepackage[utf8]{inputenc}
\usepackage[colorinlistoftodos, color=green!40, prependcaption]{todonotes}

\usepackage{hyperref}
\usepackage{amsmath}

\usepackage{mathtools}
\usepackage{physics}
\usepackage{xcolor}
\usepackage{graphicx}
\usepackage[left=23mm,right=13mm,top=35mm,columnsep=15pt]{geometry} 
\usepackage{adjustbox}
\usepackage{placeins}
\usepackage[T1]{fontenc}
\usepackage{lipsum}
\usepackage{csquotes}
\usepackage{array}

\usepackage{amsthm}
\usepackage{mathtools}
\usepackage{physics}
\usepackage{xcolor}
\usepackage{graphicx}
\usepackage[left=23mm,right=13mm,top=35mm,columnsep=15pt]{geometry} 
\usepackage{adjustbox}
\usepackage{placeins}
\usepackage[T1]{fontenc}
\usepackage{lipsum}
\usepackage{csquotes}

\begin{document}

\title{Linear optical quantum computing with a hybrid squeezed cat code}

\author{Shohei Kiryu,$^{1}$ Kosuke Fukui,$^2$ Atsushi Okamoto,$^{1}$  and Akihisa Tomita$^{1}$}
\email{tomita@ist.hokudai.ac.jp}
\affiliation{$^{1}$Graduate School of Information Science and Technology, Hokkaido University Kita14-Nishi9, Kita-ku, Sapporo 060-0814, Japan\\
$^{2}$Department of Applied Physics, School of Engineering, The University of Tokyo, 7-3-1 Hongo, Bunkyo-ku, Tokyo 113-8656, Japan}

\date{\today} 

\begin{abstract}
In recent years, squeezed cat codes with resilience to specific types of loss have been proposed as a step toward realizing fault-tolerant optical quantum computers. However, error correction for squeezed cat codes requires a strong nonlinearity, which makes its implementation challenging with current technology.
We propose a novel hybrid code that combines the squeezed cat code and the polarization qubit.
First, we propose a generation method and a universal gate set that can be implemented with a linear optical system. Then, we show the superiority of the hybrid squeezed cat code over the hybrid cat code and the squeezed cat code through numerical simulations. These results demonstrate that the hybrid squeezed cat code is a promising candidate as a new resource for optical quantum information processing.
\end{abstract}

\maketitle

\section{INTRODUCTION}

    Optical quantum computing is a promising technology for practical quantum computers, because of its room-temperature operation and compatibility with photonic communication.
    To realize optical quantum computers, a main issue is to protect the quantum states from errors due to photon loss, while the performance of most other quantum computers is limited by decoherence caused by other than photon loss.
    Bosonic codes have been reported to exhibit resilience against specific types of loss \cite{bosonic_loss1,bosonic_loss2}. 
    Bosonic codes, such as the cat code \cite{cat,cat1,cat2}, the Gottesman-Kitaev-Preskill (GKP) code \cite{GKP}, and the squeezed cat code \cite{SC, SC_KL, SC_dissipatively, SC_projective}, have been widely studied for the realization of fault-tolerant optical quantum computers \cite{myers2011coherent,bourassa2021blueprint,PRXQuantum3030309,fukui2023high}. 

    In particular, the GKP code exhibits high resistance to white noise and photon loss \cite{GKP_Analog, GKP_photonloss}, enabling universal fault-tolerant quantum computation using only linear optical operations \cite{GKP_Gaussian}. 
    Furthermore, its implementation in optical systems has been reported \cite{GKP_imp}. A generation method called the cat breeding protocol \cite{GKP_generate} was used for the GKP state implementation experiment. This method can generate GKP states using only linear optical operations. However, it requires multiple squeezed cat states as initial states, which complicates the generation process.

    The squeezed cat code is another bosonic code, based on superpositions of coherent squeezed states, and it has been suggested to offer resistance to dephasing and photon loss \cite{SC_KL}. 
    However, the existing methods for correcting errors in the squeezed cat code require strong nonlinearity, making them unfeasible with current technology.

    In recent years, hybrid approaches that combine single-photon qubit or polarization qubit with these bosonic codes have been proposed to leverage the strengths of each encoding scheme \cite{H-cat1,H-cat2,PhysRevLett125060501,H-cat_FTQC,H-cat3,fukui2024resource}. 
    For example, it has been reported that the hybridization of polarization qubits and cat codes compensates for their shortcomings, such as the probabilistic entanglement generation in polarization qubits and the probabilistic single-mode gate operation in cat codes \cite{H-cat2}.
    The hybrid cat code enables near-deterministic Bell measurements, though this is only achievable when the average photon number is large.
    Furthermore, fault-tolerant quantum computation methods using hybrid cat codes have been proposed, which offer advantages in terms of resource efficiency and photon loss tolerance \cite{H-cat_FTQC}. 
    Recently, various continuous-variable and discrete-variable hybrid protocols have also been proposed, not only for quantum computing but also for quantum information processing such as quantum communication \cite{H-protocol1,H-protocol2,H-protocol3_QKD,H-protocol4_repeater,H-protocol6,H-protocol5,H-protocol7}.

    In this paper, we propose a novel hybrid squeezed cat (H-SC) code that hybridizes the squeezed cat state with the polarization qubits, and compare it with 
    hybrid cat codes and squeezed cat codes. It was found that the H-SC code can be generated with a higher probability than the hybrid cat code. 
    With respect to gate operations, the H-SC codes were found to be easier to implement gate operations than squeezed cat codes and to have a higher probability of gate operations than the hybrid cat codes.

    Furthermore, we propose a method to compensate for a loss caused by interactions with the environment by quantum teleportation in squeezed cat codes and H-SC codes. Using this method, the H-SC code can compensate a loss with a higher probability than the squeezed cat code.

    These findings demonstrate the potential of the H-SC code as a novel and versatile resource in optical quantum information processing, paving the way for improved performance and scalability of quantum technologies.
\section{Squeezed cat code} \label{sec:develop}
From a coherent state $\ket{\alpha}$, the cat code is defined as follows
\begin{equation}
    \ket{\mathcal{C}^\pm_\alpha}=\frac{1}{\mathcal{N}^\pm_\alpha}(\ket{\alpha} \pm \ket{-\alpha}),
\end{equation}
where the normalization constants $\mathcal{N}^\pm_\alpha$ are given by
\begin{equation}
    \mathcal{N}^\pm_\alpha = \sqrt{2(1\pm e^{-2|\alpha|^2})}.
\end{equation}
In the extreme limit of high displacement ($|\alpha|^2 >>1 $), the cat code protects the encoded quantum information from amplitude damping. 
On the other hand, single-photon loss causes phase flip:
\begin{equation}
    \hat{a}\ket{\mathcal{C}^\pm_\alpha}=\frac{\alpha}{\mathcal{N}^\pm_\alpha}(\ket{\alpha} \mp \ket{-\alpha})
    \label{eq:loss_cat}
\end{equation}
As a result of the transformation, the erroneous state given by Eq. (\ref{eq:loss_cat}) remains entirely within the code space and is uncorrectable.
This problem can be solved by applying a squeezing operation to the cat code as described below.

Squeezing operation to vacuum followed by displacement operation results in a squeezed coherent state:
\begin{equation*}
    \ket{\alpha,\xi}=\hat{D}(\alpha)\hat{S}(\xi)\ket{0},
\end{equation*}
where $\hat{D}(\alpha)$ and $\hat{S}(\xi)$ are the displacement operator and the squeezing operator, respectively, defined by
\begin{equation*}
    \hat{D}(\alpha) = e^{\alpha\hat{a}^\dagger-\alpha^* \hat{a}}
\end{equation*}
and
\begin{equation*}
    \hat{S}(\xi) = e^{\frac{1}{2}(\xi^* \hat{a}^2 - \xi(\hat{a}^\dagger)^2)}.
\end{equation*} 

The squeezed cat code is defined as
\begin{equation}
    \ket{\mathcal{C}^\pm_{\alpha,\xi}}=\frac{1}{\mathcal{N}^\pm_{\alpha,\xi}}(\ket{\alpha,\xi} \pm \ket{-\alpha,\xi}) .
\end{equation}
The normalization constant $\mathcal{N}^\pm_{\alpha,\xi}$ is given by
\begin{equation*}
    \mathcal{N}^\pm_{\alpha,\xi} = \sqrt{2\left(1\pm e^{-2|\alpha \cosh(r)+e^{i\theta}\alpha^* \sinh(r)|^2}\right)},
\end{equation*}
with the complex squeezing parameter defined as $\xi = re^{i\theta}$.
When the single-photon loss $\hat{a}$ occurs in this squeezed cat code, the change of the state can be calculated using the transformation relations of the annihilation operator by a displacement operator and a squeezing operator, 
\begin{align*}
    \hat{D}^\dagger(\alpha) \hat{a}\hat{D}(\alpha) &=  \hat{a}+\alpha , \\
    \hat{S}^\dagger (\xi)\hat{a}\hat{S}(\xi) &= \hat{a} \cosh r - \hat{a}^\dagger e^{i \theta} \sinh r.
\end{align*}
The resulting state becomes 
\begin{align}\label{eq:loss_squeezed_cat}
    \hat{a}\ket{\mathcal{C}^\pm_{\alpha,\xi}} =& \frac{\alpha}{ \mathcal{N}^\pm_{\alpha,\xi}} \left(\hat{D}(\alpha) \mp \hat{D}(-\alpha) \right) \hat{S}(\xi) \ket{0} \nonumber\\
    &- \frac{e^{2i \theta}\sinh r}{ \mathcal{N}^\pm_{\alpha,\xi}} \left(\hat{D}(\alpha) \pm \hat{D}(-\alpha) \right) \hat{S}(\xi) \hat{a}^\dagger \ket{0} \nonumber\\
    =& c\ket{\mathcal{C}^\mp_{\alpha,\xi}} + d\ket{\tilde{\mathcal{C}}^\pm_{\alpha,\xi}} .
\end{align}
Here, $c$ and $d$ are constants that depend on the squeezing parameter $\xi$ and the amplitude $\alpha$ of the squeezed cat code. It is notable that $\ket{\tilde{\mathcal{C}}^\mp_{\alpha,\xi}}$ is a state that does not belong to the code space. In other words, the state $\hat{a}\ket{\mathcal{C}^\pm_{\alpha,\xi}}$ after the photon loss spans both the code space and the orthogonal error space. Since the squeezed cat code is not an eigenstate of $\hat{a}$, a complete bit-flip does not occur even when the single-photon loss occurs. For this reason, the squeezed cat code is resistant to the single-photon loss. Furthermore, the Knill-Laflamme error correction conditions suggest that the code may have resistance to both phase damping and photon loss \cite{SC_KL}. However, no operation has been proposed to restore the loss state (\ref{eq:loss_squeezed_cat}) to its original state in optical systems. To solve the optical implementation issue, we propose a hybrid code that combines the squeezed cat code and the polarization qubits. We will show that the photon loss can be compensated by a practical optical system in sec.VI.
\section{Hybrid Squeezed Cat code} \label{sec:develop2}
Before introducing the hybrid squeezed cat code, we consider the loss tolerance of the hybrid code. We take the hybrid cat (H-cat) code as an example. The H-cat code \cite{H-cat1,PhysRevLett125060501,H-cat_FTQC}, which hybridizes the cat code and a polarization qubit, is defined as
\begin{equation}
    \ket{0_\textrm{L}} = \ket{+}\ket{\mathcal{C}^+_\alpha} ,
\end{equation}
\begin{equation}
    \ket{1_\textrm{L}} = \ket{-}\ket{\mathcal{C}^-_\alpha}.
\end{equation}
Here, $\ket{\pm}$ represent the superposition states of single-photon polarization states $\ket{H}$ and $\ket{V}$: 
\begin{equation*}
    \ket{\pm} = \frac{1}{\sqrt{2}}(\ket{H} \pm \ket{V}),
\end{equation*}
where $\ket{H}$ and $\ket{V}$ are horizontal and vertical polarization states. Unlike the cat code, the H-cat code can detect photon loss. When a photon is lost in the cat state of the H-cat code, it behaves as  
\begin{equation*}
    \hat{a}\ket{+}\ket{\mathcal{C}^+_\alpha} \propto \ket{+}\ket{\mathcal{C}^-_\alpha} \neq \ket{-}\ket{\mathcal{C}^-_\alpha}.
\end{equation*}
Since the polarization qubit remains unchanged despite the photon loss in the cat state, the orthogonality between the polarization qubits $\ket{+}$ and $\ket{-}$ allows error detection. In other words, H-cat codes are error correctable as long as the polarization qubit is maintained even in the presence of single-photon loss to the cat state. However, if the photon of a polarization qubit is also lost, it cannot be corrected.
Therefore, it is necessary to construct a hybrid code using bosonic codes that are resilient to photon loss.
We propose a hybrid code of squeezed cat states and polarization qubits. 
The hybrid squeezed cat (H-SC) code is defined as 
\begin{equation}
    \ket{0_L} = \ket{+}\ket{\mathcal{C}^+_{\alpha,\xi}},
\end{equation}
\begin{equation}
    \ket{1_L} = \ket{-}\ket{\mathcal{C}^-_{\alpha,\xi}}.
\end{equation}
As discussed in the previous section, squeezed cat codes can detect errors even in the presence of loss; therefore, it is expected that the single-photon losses occurring in both polarization qubits and SC states can also be detected.
However, for the proposed code to be feasible, it must be able to perform universal gate operations. We will discuss these issues in detail in Sec.IV and Sec.V.
\section{Generation of entanglement state} \label{sec:develop3}
When performing quantum computation using H-SC codes, as described in detail in Sec. V, the following hybrid entangled states are required as ancilla qubits:
\begin{equation} 
    \ket{\psi} = \frac{1}{\sqrt{2}}\left( \ket{H}\ket{\alpha_f,\xi} + \ket{V}\ket{-\alpha_f,\xi} \right).
    \label{eq:Hybrid_ent_state}
\end{equation} 
It is known that such a state can be generated by a weak cross-Kerr nonlinear interaction when $\xi=0$ \cite{gane_H-cat1,gane_H-cat2}. However, realizing an ideal nonlinear interaction with current technology is highly challenging. A probabilistic generation scheme using linear optics and photon number detection has been proposed \cite{gene_Hybrid}. In this paper, we propose an optical system for generating H-SC entangled states by extending this scheme.

The proposed scheme starts with a squeezed cat state $\ket{\textrm{SCS}(\alpha_i,\xi) }$, a squeezed vacuum state $\ket{0,\xi}$, and a maximally entangled polarization qubit $\ket{\Psi^{+}}=(\ket{H}\ket{V}+\ket{V}\ket{H})/\sqrt{2}$. The squeezed cat state with an amplitude $\alpha_i$ is defined as follows:
\begin{equation} 
    \ket{\textrm{SCS}(\alpha_i,\xi) }=\ket{\mathcal{C}^+_{\alpha_i,\xi}}= \frac{1}{\mathcal{N}^+_{\alpha_i,\xi}} (\ket{\alpha_i,\xi} + \ket{-\alpha_i,\xi}) .
\end{equation} 
The hybrid entangled state as expressed in Eq.(\ref{eq:Hybrid_ent_state}) can be generated from the initial states by a circuit consisting of a displacement operation, a variable beam splitter, a half beam splitter(HBS), polarization beam splitters(PBSs), and photon number resolving detectors, as shown in Fig.1.
\begin{figure}[t] 
    \centering \includegraphics[width=0.45\textwidth]{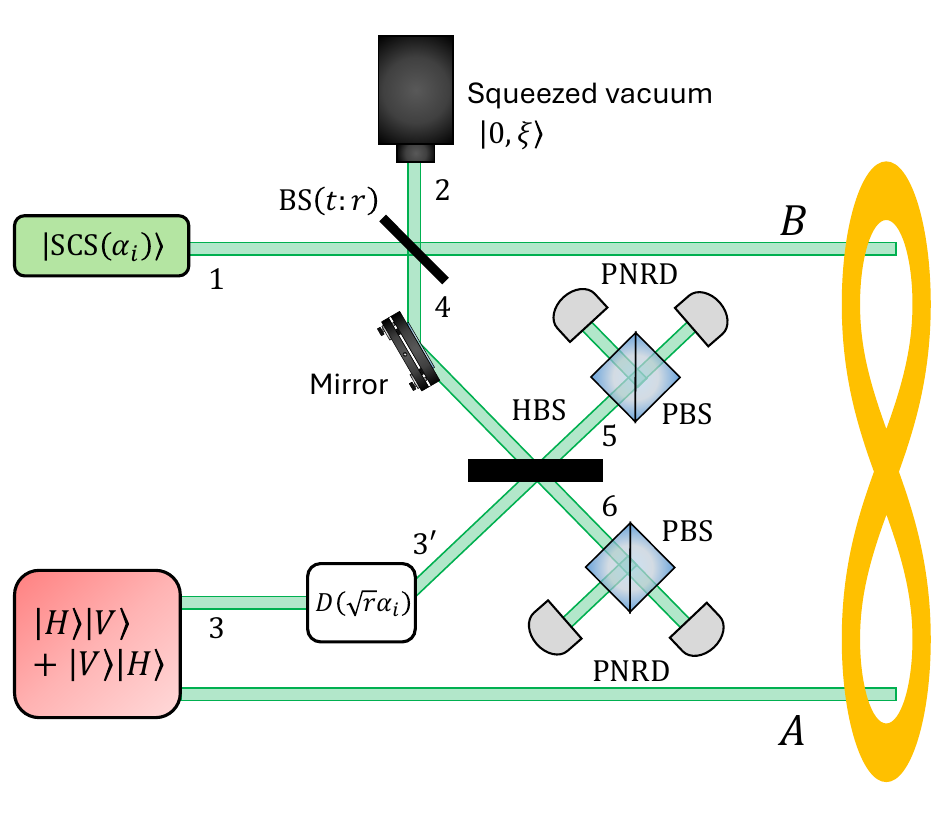} \caption{The generation circuit of the hybrid entangled states. It consists of a displacement operation, a variable beam splitter, a half beam splitter (HBS), polarization beam splitters (PBSs), and photon number resolving detectors. Nonlinear transformations are unnecessary.} 
\end{figure} 
The input squeezed coherent state and the squeezed vacuum state are transformed by the variable beam splitter BS($t:r$) with transmittance $t$ and reflectance $r$ $(=1-t)$ to 
\begin{equation} 
\ket{\alpha,\xi}_1 \ket{0,\xi}_2 \to \ket{\sqrt{t}\alpha,\xi}_B \ket{\sqrt{r}\alpha,\xi}_4 . 
\end{equation} 
The input to one port (4) of the HBS is given by the superposition of the following states:
\begin{equation*}
    \ket{\pm \sqrt{r} \alpha_i,\xi}_4=\hat{D}_4(\pm \sqrt{r} \alpha_i) \hat{S}_4(\xi)\ket{0},
\end{equation*}
while the input to the other port (3') is given by
\begin{equation*}
\begin{split}
    &\hat{D}_3(\sqrt{r} \alpha_i)\ket{\Psi^{+}}\\
    &=\frac{1}{\sqrt{2}}\left(\hat{D}_3(\sqrt{r} \alpha_i)\ket{H}_3 \ket{V}_A + \hat{D}_3(\sqrt{r} \alpha_i) \ket{V}_3 \ket{H}_A \right).
\end{split}
\end{equation*}
Therefore, the effect of the HBS can be expressed as
\begin{equation}
\begin{split}
    \hat{D}_4(\alpha)\hat{D}_{3'}(\beta)\hat{S}_4(\xi) 
    \to \hat{D}_{5}\left(\frac{-\alpha+\beta}{\sqrt{2}}\right)\hat{D}_{6}\left(\frac{\alpha+\beta}{\sqrt{2}}\right)\hat{S}_{56},
\end{split}
\end{equation}
where the operator $\hat{S}_{56}$ denotes 
\begin{equation*}
    \hat{S}_{56}=\hat{S}_{5}\left(\frac{\xi}{2}\right)\hat{S}_{6}\left(\frac{\xi}{2}\right)\hat{S}^{(2)}_{56}\left(\frac{\xi}{2}\right) .
\end{equation*}
A two-mode squeezing operator $\hat{S}^{(2)}_{56}\left(\frac{\xi}{2}\right)$ is defined as follows:
\begin{equation*}
    \hat{S}^{(2)}_{56}\left(\frac{\xi}{2}\right) =  e^{\frac{1}{2}(\xi^* \hat{a}_5\hat{a}_6 - \xi\hat{a}_5^\dagger \hat{a}_6^\dagger)}.
\end{equation*}
These transformations result in the state before measurement as 
\begin{equation}
\label{bf_measure}
    \begin{split}
    &\ket{\psi_{\phi}} = \frac{1}{2 \mathcal{N}^+_{\alpha,\xi}}\\
    &\Big[ 
    \ket{H}_{A}\hat{D}_{6}(\sqrt{2r}\alpha_i)\hat{S}_{56}(\ket{V}_5 \ket{0}_6 + \ket{0}_5 \ket{V}_6)\ket{\sqrt{t}\alpha_i,\xi}_B \\
    &+\ket{H}_{A}\hat{D}_{5}(\sqrt{2r}\alpha_i)\hat{S}_{56}(\ket{V}_5 \ket{0}_6 + \ket{0}_5 \ket{V}_6)\ket{-\sqrt{t}\alpha_i,\xi}_B  \\
    &+\ket{V}_{A}\hat{D}_{6}(\sqrt{2r}\alpha_i)\hat{S}_{56}(\ket{H}_5 \ket{0}_6 + \ket{0}_5 \ket{H}_6)\ket{\sqrt{t}\alpha_i,\xi}_B \\
    &+\ket{V}_{A}\hat{D}_{5}(\sqrt{2r}\alpha_i)\hat{S}_{56}(\ket{H}_5 \ket{0}_6 + \ket{0}_5 \ket{H}_6)\ket{-\sqrt{t}\alpha_i,\xi}_B
    \Big].
    \end{split}
\end{equation}
Applying the following measurement operator to the state $\ket{\psi_{\phi}}$
\begin{equation*}
    \Pi = I_A \otimes \ket{0}\bra{0}_{5H}\otimes \ket{1}\bra{1}_{5V}\otimes \ket{1}\bra{1}_{6H}\otimes \ket{0}\bra{0}_{6V}\otimes I_B ,
\end{equation*}
we obtain a hybrid entangled state as
\begin{align*}
    \rho &= \frac{Tr_{56}(\Pi \ket{\psi_{\phi}}\bra{\psi_{\phi}})}{\bra{\psi_{\phi}} \Pi\ket{\psi_{\phi}}} \\
    &\propto (\ket{H}_A\ket{\alpha_f,\xi}_B + \ket{V}_A\ket{-\alpha_f,\xi}_B)\\
    & \ \ ({}_A\bra{H}_B\bra{\alpha_f,\xi} + {}_A\bra{V}_B\bra{-\alpha_f,\xi}) ,
\end{align*}
where the amplitude of the generated state is $\alpha_f = \sqrt{t}\alpha_i$.

 The success probability of generating the hybrid entanglement state is given by
\begin{equation}
    P^{\phi} = \bra{\psi_{\phi}} \Pi\ket{\psi_{\phi}} .
    \label{eq:generating_rate}
\end{equation}
The explicit form of the success probability is complicated and difficult to interpret. However, in the special case when $\xi = 0$, Eq.(\ref{eq:generating_rate}) can be written as 
\begin{equation*} 
    P^{\phi} = \left(\frac{1}{\mathcal{N}^+_{\alpha_i,0}} \right)^2 (1-t)\alpha_i^2 e^{-2(1-t)\alpha^2_i},
\end{equation*}
which takes the maximum value at $(1-t)\alpha_i^2=1/2$.

The success probability can be 
increased by utilizing the states generated from different measurement outcomes as well, because the hybrid entanglement state can also be produced using the following measurement operator 
\begin{equation*}
    \Pi' = I_A \otimes \ket{1}\bra{1}_{5H}\otimes \ket{0}\bra{0}_{5V}\otimes \ket{0}\bra{0}_{6H}\otimes \ket{1}\bra{1}_{6V}\otimes I_B  .
\end{equation*}
Compared to the measurement operator $\Pi$, $\ket{0}$ and $\ket{1}$ are interchanged in the measurement operator $\Pi'$. Therefore, if the measurement outcome is $\Pi'$, the desired state can be obtained by applying a bit flip on the polarization qubit or a $\pi$ phase shift on the squeezed cat state.
The generation probability of the H-SC entangled states with the measurement operator $\Pi'$ is also $P^{\phi}$. By applying the bit flip (or $\pi$ phase shift) based on the measurement results, we can double the probability of success.

The success probability of state generation in the circuit shown in Fig. 1. depends on the transmittance of the variable beam splitter $t$, the amplitude $\alpha_i$, and the squeezing parameter $\xi$ of the initial squeezed cat state. The results of a numerical simulation of the state generation probability are shown in Fig.2. Note that the average photon number differs between a coherent state $\ket{\alpha}$ and a coherent squeezed state $\ket{\alpha,\xi}$ with the same amplitude. Therefore, to ensure a fair comparison, the average photon number of the squeezed cat state was fixed at $\bar{n} = 2.0$.

\begin{figure}[t] \centering \includegraphics[width=0.45\textwidth]{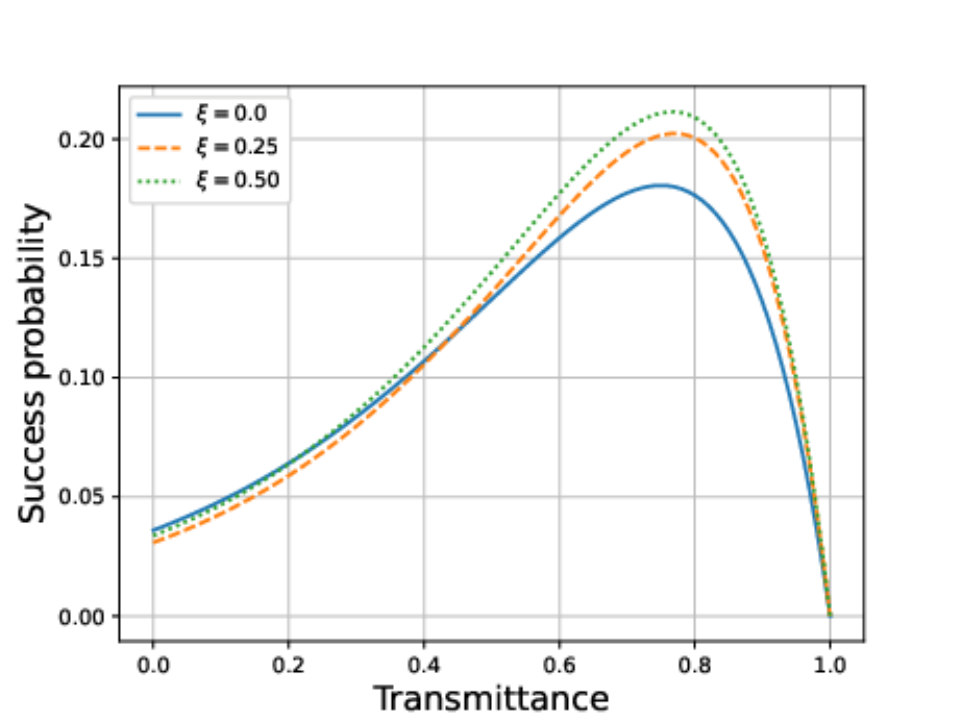} \caption{Success probability of generating the H-SC code. The blue (solid) line shows the success probability for a hybrid cat state with squeezing parameter $\xi = 0$. The orange (broken line) and green (dots) represent the success probabilities for H-SC states with $\xi = 0.25$ and $\xi = 0.50$, respectively. For $\xi = 0.25$ and $0.50$, the corresponding amplitudes are approximately $|\alpha| \approx 1.39$ and $1.32$, respectively.} \end{figure}

The results of numerical simulations indicate that in the region of high transmittance $t$, the success probability increases with squeezing. However, in the region of low transmittance $t$, the squeezing does not necessarily result in a higher generation probability. 
Furthermore, since squeezing itself requires additional resources, we need to carefully consider the advantage of the H-SC code, if the generation costs matter \cite{PhysRevA76060301}. 

\section{Universal Gate Set} \label{sec:develop4}

In this section, we discuss gate operations for the H-SC code. In quantum computation, a universal gate set is required to perform an arbitrary operation. This set typically includes bit flip (X) gate, $\theta$-rotation gate ($Z(\theta)$), Hadamard (H) gate, and Controlled-NOT (CNOT) gate. In the following, we show that all these operations can be implemented with the H-SC code.

The X gate can be implemented by rotating the polarization qubit by $\pi / 2$ and applying $i\hat{D}(i\frac{\pi}{4\xi})$ to the squeezed cat state. The $Z(\theta)$ gate is executed by applying the transformation $\ket{-} \to e^{i\theta} \ket{-}$ to the polarization qubit. These operations can be implemented using wave plates for polarization qubit operations and optical devices such as modulators for displacement operations. 
In the case of non-hybrid squeezed cat code, the $Z(\theta)$ gate is implemented by the operation $e^{i\theta\hat{a}^\dagger\hat{a}}$, which implies phase rotation of $\theta$ per photon. This operation requires a significant optical non-linearity. Therefore, the $Z(\theta)$ gate for the H-SC code is easier to implement than for the SC code. 
In this regard, the H-cat code can be implemented similarly to the H-SC code and offers the same advantages
\cite{H-cat2}.

The H and CNOT gate can be implemented by gate teleportation, where ancilla qubits $\ket{\psi_H} \propto \ket{0_L, 0_L} + \ket{0_L, 1_L} + \ket{1_L, 0_L} - \ket{1_L, 1_L}$ for H gate and $\ket{\psi_{CNOT}} \propto \ket{0_L, 0_L, 0_L, 0_L} + \ket{0_L, 0_L, 1_L, 1_L} + \ket{1_L, 1_L, 0_L, 0_L} - \ket{1_L, 1_L, 1_L, 1_L}$ for CNOT gate are prepared. Entanglement generation and Bell measurements are used to execute the gates. These ancilla qubits can be generated from multiple hybrid entanglement states of Eq.(\ref{eq:Hybrid_ent_state}) with the HBS and Bell measurements \cite{H-cat2}.

The circuit for operating an H gate on a state $\ket{\psi}$ using gate teleportation is shown in Fig. 3.
To apply the H gate to the state $\ket{\psi}$, one needs to perform a Bell measurement on the state $\ket{\psi}$ and the first qubit of the ancilla qubits $\ket{\psi_H}$. The Bell measurement can be implemented by HBS, PBS, photon number resolving detector (PNRD), and photon detector (PD).
Based on the result of the Bell measurement, a unitary transformation $\hat{U}$ is applied to the second qubit of the ancilla qubits. The operation $\hat{U}$ can be implemented with wave plates and displacement operations. The output state then becomes $\hat{H}\ket{\psi}$.

\begin{figure}[t] \centering \includegraphics[width=0.45\textwidth]{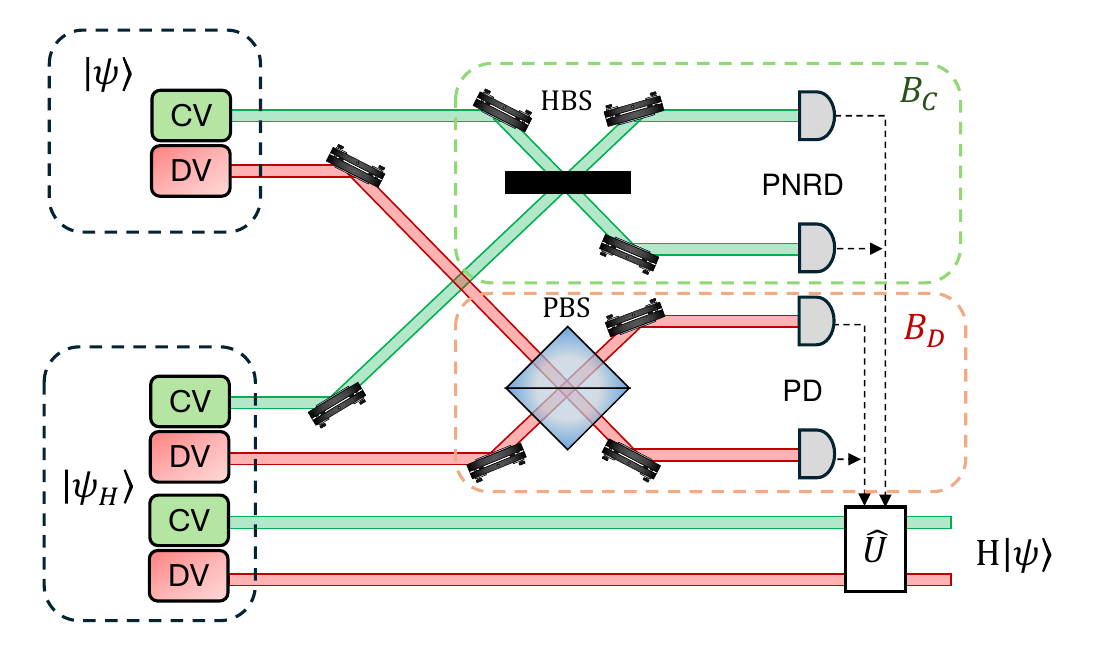} \caption{Optical system for performing the H gate on the H-SC code. $B_D$ is the Bell measurement for the polarization qubit, and $B_C$ is the Bell measurement for the squeezed cat state. PNRDs represent photon number resolving detectors, and PDs represent photon detectors. The operation $\hat{U}$ can be implemented with wave plates and displacement operations.} \end{figure}

The H gate and CNOT gate operate probabilistically, with their success probabilities determined by the success probability of Bell measurements. 
Note that the Bell measurement on the hybrid code is successful if a measurement succeeds on either the polarization qubit or the squeezed cat state, as shown in the table \ref{Table:HBM_result} below.  
\begin{table}[h]
 \begin{center}
   \caption{Hybrid Bell measurement results}
   \label{Table:HBM_result}
        \begin{tabular}{>{\centering\arraybackslash}p{2cm}>{\centering\arraybackslash}p{2cm}>{\centering\arraybackslash}p{2cm}} \hline
            $B_D$ & $B_C$ & Result \\ \hline
            Success  & Success  & Success  \\
            Failure  & Success  & Success  \\
            Success  & Failure  & Success  \\
            Failure & Failure & Failure \\ \hline
        \end{tabular}
    \end{center}
\end{table}
Here, $B_D$ represents the Bell measurement of the polarization qubit, and $B_C$ represents the Bell measurement of the squeezed cat state.

The Bell measurement for the polarization qubit $B_C$ succeeds only probabilistically with the probability at most  \(1/2\) using linear optics. We will use this maximal value hereafter.
In contrast, the Bell measurement for the squeezed cat state $B_D$ can be nearly deterministic if the average photon number is large. 
The Bell measurement is successful if the photon number measurement yields an outcome other than zero, and the corresponding squeezed vacuum state $\ket{0,\xi}$ collapses to $\ket{0}$. 
%
Therefore, the success probability for the Bell measurement on squeezed cat states is expressed as
\[
P_{\text{CV}}=\left(1 - \left| \braket{0 }{\mathcal{C}^+_{\sqrt{2}\alpha,\xi}} \right|^2 \right) \left| \braket{0 }{ 0, \xi} \right|^2.
\]
The success probability depends on the squeezing parameter $\xi$ and the amplitude $\alpha$. 
The total success probability of the hybrid Bell measurement then becomes
\begin{equation}
P_{\text{hybrid}} = 1 - \frac{1}{2}(1 - P_{\text{CV}}).    
\end{equation}
The present formulation is consistent with the description in Ref.\cite{H-cat2} when $\xi = 0$.




The hybrid Bell measurement can achieve a high success probability \cite{H-cat3}.
In general, the success probability increases as both squeezing and amplitude increase. However, when the average photon number $\bar{n}$ is fixed, the success probability of the Bell measurement is maximized by selecting the appropriate squeezing parameter.
Figure 4 shows the results of numerical simulations on the optimal squeezing parameter and the corresponding success probability of the hybrid Bell measurement as a function of the average photon number.
As the average photon number $\bar{n}$ increases, the optimal squeezing parameter also increases for small average photon numbers. However, we observed that, beyond an average photon number of around $\bar{n} = 0.7$, the optimal squeezing parameter begins to decrease. 
Two main factors, which are in a trade-off relationship, cause the failure of Bell measurements on the squeezed cat state. 
The first is the indistinguishability between the coherent squeezed state $\ket{\alpha,\xi}$ and the vacuum state $\ket{0}$. The second is the low probability of detecting zero photon when performing a photon number measurement on a squeezed vacuum state $\ket{0,\xi}$ that has the same squeezing parameter as the squeezed cat state.

\begin{figure}[t] \centering \includegraphics[width=0.45\textwidth]{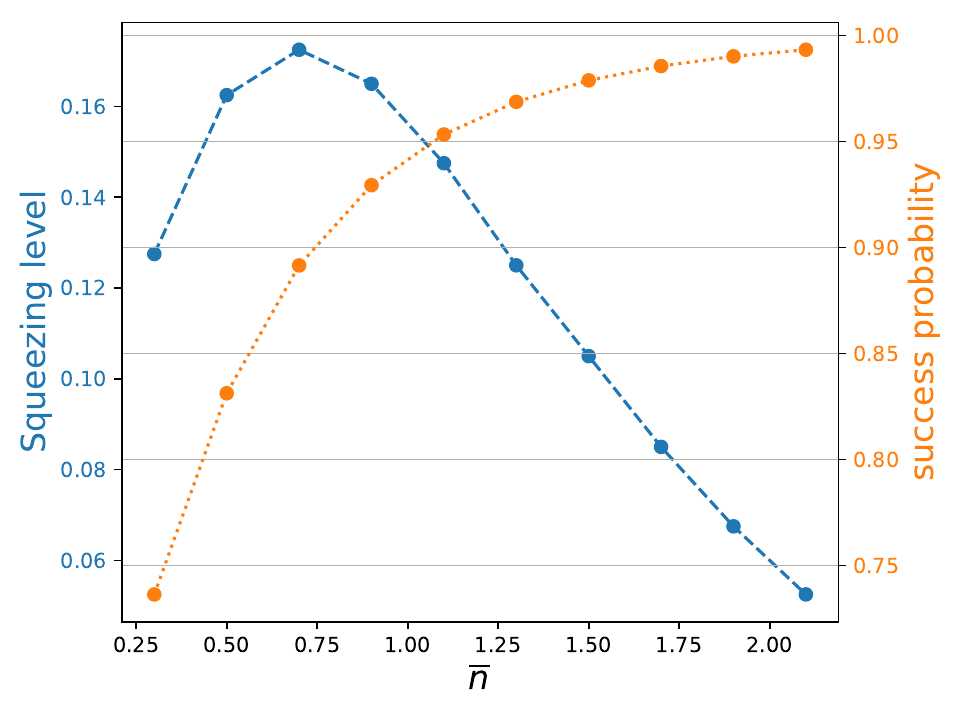} \caption{Optimal squeezing parameter for maximizing the success probability of hybrid Bell measurements, and the corresponding success probability. The blue (broken) line represents the optimal squeezing parameter, and the orange (dots) represents the success probability of the hybrid Bell measurement. 
The optimal squeezing parameter takes the maximum value 
$\xi \approx 0.17$ at $\bar{n} = 0.7$, where the corresponding amplitude is approximately $\alpha \approx 2$.} 
\end{figure}
A higher squeezing parameter $\xi$ renders the squeezed state more distinct from the vacuum state and reduces the failure probability due to the first factor. 
However, the high squeezing parameter increases the failure probability due to the second factor, because the average photon number of the squeezed vacuum rises. When the squeezing parameter is low, the squeezed vacuum state $\ket{0,\xi}$ is nearly indistinguishable from the vacuum state $\ket{0}$, which makes the probability of detecting zero photons in the photon number measurement high. Conversely, as the squeezing parameter increases, the probability of detecting other even-photon numbers increases.
Thus, there exists an optimal squeezing parameter that minimizes the failure probabilities caused by the two factors and yields the highest success probability for the hybrid Bell measurement.

\section{loss compensation} \label{sec:develop5}

In optical platforms, photon loss is the dominant source of decoherence. Photon loss can be modeled by applying a beam splitter to the input state $\rho$ and a vacuum state, followed by tracing out the environmental degrees of freedom \cite{Kim1995Phase}. This process can also be described using a Kraus representation of the loss map, as shown below.
\begin{equation}\label{eq:loss}
\Lambda_\gamma [\rho] = \sum^{\infty}_{j=0} K_j (\gamma) \rho K_j ^\dagger(\gamma).
\end{equation}
Here, the Kraus operators are given by
\begin{equation}
K_j (\gamma)= \sqrt{\frac{\gamma^j}{j!}}(1-\gamma)^\frac{\hat{n}}{2}\hat{a}^j = \sqrt{\frac{1}{j!}\left( \frac{\gamma}{1-\gamma}\right)^j}\hat{a}^j (1-\gamma)^\frac{\hat{n}}{2},
\end{equation}
where $\hat{a}$ is the annihilation operator, and $\gamma$ is loss parameters.
Since this channel is trace-preserving, the Kraus operators satisfy the completeness relation $\sum_j K_j^\dagger (\gamma) K_j (\gamma)=\mathbb{I}$.

\begin{figure}[t] \centering \includegraphics[width=0.45\textwidth]{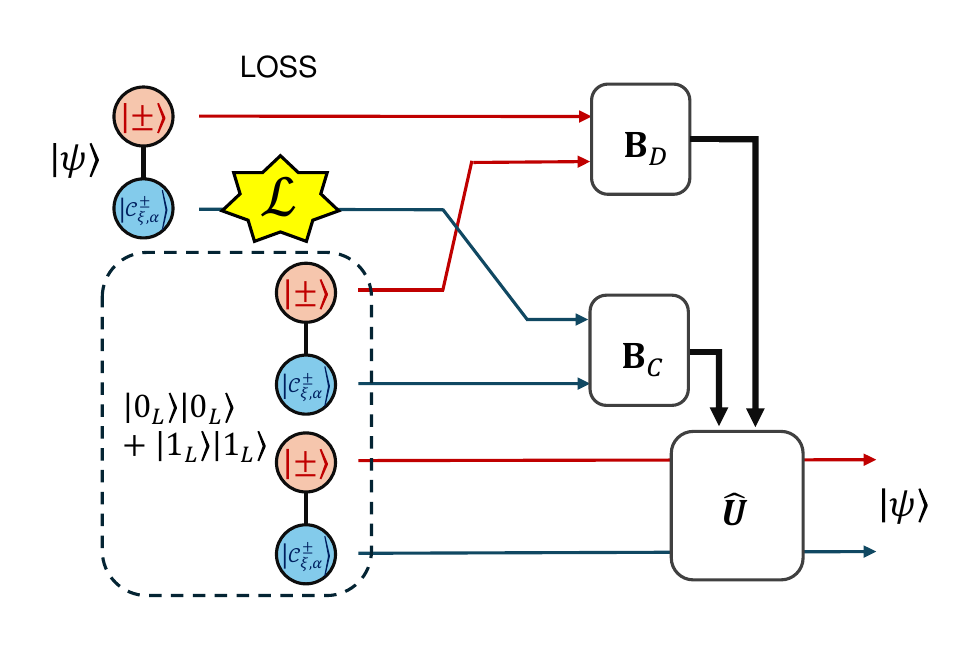} \caption{Loss compensation circuit through quantum teleportation. $B_D$ represents the Bell measurement of the polarization qubit, and $B_C$ represents the Bell measurement of the squeezed cat state. The ancilla qubits required are $(\ket{0_L}\ket{0_L} + \ket{1_L}\ket{1_L})/\sqrt{2}$. Losses are assumed not to occur in the ancillary qubits.} 
\end{figure}

The hybrid teleportation can compensate for the loss in the H-SC code, allowing the state $\ket{\psi}$ to be restored, as shown in Fig.5.
As long as the hybrid Bell measurement with the ancilla qubits correctly identifies the Bell state, one can restore the lossless state $\ket{\psi}$ by applying a unitary transformation to the ancilla qubit based on the measurement outcomes. 
The ancilla qubits $\ket{\Phi^+} = (\ket{0_L}\ket{0_L} + \ket{1_L}\ket{1_L})/\sqrt{2}$ are needed to compensate for the loss.
Furthermore, it is assumed that there are no losses or measurement errors in the ancilla qubits. 

The operation of the loss compensation circuit shown in Fig. 5 can be analytically described as follows.
Let the input state after the loss be denoted as $\rho_L = \Lambda_\gamma[\ket{\psi}\bra{\psi}]$, where $\Lambda_\gamma$ represents the lossy channel defined by Eq. (\ref{eq:loss}).
The joint state becomes $\rho_{\text{joint}} = \rho_L \otimes \ket{\Phi^+}\bra{\Phi^+}$.

After performing a Bell measurement $\mathcal{B}$ on the lossy state and one qubit of the ancilla, the state collapses onto one of the Bell basis.
The second ancilla qubit (not measured) then becomes $U_i \ket{\phi}$, 
where $U_i$ is a Pauli operator depending on the Bell measurement outcome $i \in \{00,01,10,11\}$.

The final corrected state is given by
\[
    \rho_{\text{out}} = U_i^\dagger \rho_i U_i,
\]
where $\rho_i$ is the post-measurement state conditioned on outcome $i$, and $U_i$ is the correction operation.
The success of the loss compensation depends on the correct identification of the Bell state in the hybrid Bell measurement and teleportation.

In the following, we compare the success probability of the loss compensation for the H-SC state with that for the SC state, which can be compensated for the loss using a similar quantum circuit.
The results of the numerical simulations are shown in Fig. 6. The squeezing parameter of the states was chosen to maximize the success probability. It was found that the optimal squeezing parameter is $\xi>0$ in all regions, and the protocol for H-SC scheme can be implemented with a higher success probability than the hybrid cat scheme.
The results indicate that the H-SC code can be compensated for a loss with a higher probability, particularly when the average photon number is low. As the average photon number increases, the squeezed cat code also achieves a high success probability of loss compensation.

\begin{figure}[t] \centering \includegraphics[width=0.45\textwidth]{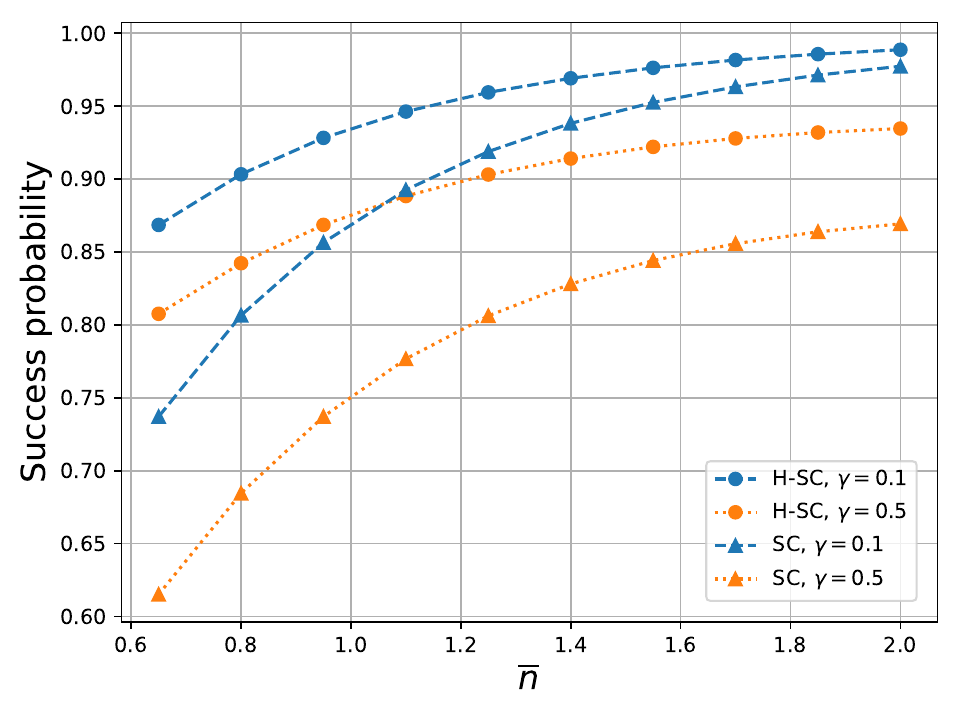} \caption{Success probability of loss compensation. The blue (broken) line represents the case with low loss, while the orange (dots) represents the case with high loss. The round markers represent the success probability of loss compensation with the H-SC code, and the triangle markers represent that with the squeezed cat code.} \end{figure}

\section{CONCLUSIONS} \label{sec:conclusions}
In this paper, we proposed a hybrid squeezed cat (H-SC) code. First, we presented its generation method and investigated its generation efficiency. We demonstrated that H-SC states can be generated without nonlinear interactions, provided that squeezed cat states, vacuum states, and entangled polarization qubits are prepared. Next, we defined the universal gate set for the H-SC code. Similarly to the generation scheme, these gate sets do not require any nonlinear interactions. This means that once the initial states necessary for the H-SC state generation are prepared, all subsequent operations can be easily implemented using current optical technology. Moreover, several gate operations can be performed more easily with the H-SC code compared to the squeezed cat code.

We then demonstrated through numerical simulations that the H-SC code achieves a gate operation with a higher success probability than conventional squeezed cat codes or hybrid cat codes. This improvement is particularly pronounced when the average photon number is low.

Finally, we proposed a method to compensate for the losses in the H-SC code and the squeezed cat code using quantum teleportation. We discovered that the H-SC code, which is redundantly encoded using polarization, has a higher probability of successfully compensating for loss compared to the squeezed cat code. However, as the average photon number increases, the squeezed cat code also has a high probability of loss compensation, indicating its potential as a resource for quantum computation in the large average photon number regime. 
These findings suggest that both the H-SC code and the SC code have potential, as the cat state used in the simulation has an average photon number achievable with current experimental techniques \cite{exp_cat5,exp_cat3,exp_cat2,exp_cat1,exp_cat4,exp_cat6}.

To utilize the H-SC code as a resource for universal fault-tolerant quantum computation, it will be necessary to concatenate it with quantum error-correcting codes, such as surface codes \cite{surface,XZZX}. In fact, it has been reported that GKP codes and cat codes improve their performance when concatenated with the surface codes \cite{XZZX_cat,surface_GKP,XZZX_GKP}. 
Additionally, the loss performance can be improved by hybridization with more complex states such as four-component squeezed cat codes \cite{PhysRevX10011058,endo2022quantum}. However, this approach involves a higher level of complexity in the generation process.

While this study focused on implementing the H-SC code with polarization qubits and squeezed cat states of light, it is possible to use other states. For example, the H-SC states can be constructed with single photon-vacuum qubits.
Furthermore, the H-SC code can be potentially implemented in various physical systems other than light. In fact, the GKP code has been experimentally demonstrated in ion traps \cite{GKP_ion} and superconducting circuit platforms \cite{GKP_sup}. The development of methods for generating the H-SC code in such physical systems, as well as gate operation techniques and loss resistance, is an important challenge for the future.

\section*{Acknowledgements} \label{sec:develop6}
This work was partly supported by JST PRESTO Grant No. JPMJPR23FA and JST Moonshot R\&D Grant No. JPMJMS2061.

\bibliography{main} 

\begin{thebibliography}{10}

\bibitem{bosonic_loss1}
B~M Terhal, J~Conrad, and C~Vuillot.
\newblock Towards scalable bosonic quantum error correction.
\newblock {\em Quantum Science and Technology}, 5(4):043001, 2020.

\bibitem{bosonic_loss2}
Atharv Joshi, Kyungjoo Noh, and Yvonne~Y Gao.
\newblock Quantum information processing with bosonic qubits in circuit {QED}.
\newblock {\em Quantum Science and Technology}, 6(3):033001, 2021.

\bibitem{cat}
P.~T. Cochrane, G.~J. Milburn, and W.~J. Munro.
\newblock Macroscopically distinct quantum-superposition states as a bosonic code for amplitude damping.
\newblock {\em Phys. Rev. A}, 59:2631--2634, 1999.

\bibitem{cat1}
H.~Jeong and M.~S. Kim.
\newblock Efficient quantum computation using coherent states.
\newblock {\em Phys. Rev. A}, 65:042305, 2002.

\bibitem{cat2}
T.~C. Ralph, A.~Gilchrist, G.~J. Milburn, W.~J. Munro, and S.~Glancy.
\newblock Quantum computation with optical coherent states.
\newblock {\em Phys. Rev. A}, 68:042319, 2003.

\bibitem{GKP}
Daniel Gottesman, Alexei Kitaev, and John Preskill.
\newblock Encoding a qubit in an oscillator.
\newblock {\em Phys. Rev. A}, 64:012310, 2001.

\bibitem{SC}
David~S. Schlegel, Fabrizio Minganti, and Vincenzo Savona.
\newblock Quantum error correction using squeezed schr\"odinger cat states.
\newblock {\em Phys. Rev. A}, 106:022431, 2022.

\bibitem{SC_KL}
Qian Xu, Guo Zheng, Yu-Xin Wang, Peter Zoller, Aashish~A. Clerk, and Liang Jiang.
\newblock Autonomous quantum error correction and fault-tolerant quantum computation with squeezed cat qubits.
\newblock {\em npj Quantum Information}, 9:78, 2023.

\bibitem{SC_dissipatively}
Timo Hillmann and Fernando Quijandr\'{\i}a.
\newblock Quantum error correction with dissipatively stabilized squeezed-cat qubits.
\newblock {\em Phys. Rev. A}, 107:032423, 2023.

\bibitem{SC_projective}
Suguru Endo, Keitaro Anai, Yuichiro Matsuzaki, Yuuki Tokunaga, and Yasunari Suzuki.
\newblock Projective squeezing for translation symmetric bosonic codes.
\newblock arXiv:2403.14218 [quant-ph], 2024.

\bibitem{myers2011coherent}
Casey~R Myers and Timothy~C Ralph.
\newblock Coherent state topological cluster state production.
\newblock {\em New Journal of Physics}, 13(11):115015, 2011.

\bibitem{bourassa2021blueprint}
J~Eli Bourassa, Rafael~N Alexander, Michael Vasmer, Ashlesha Patil, Ilan Tzitrin, Takaya Matsuura, Daiqin Su, Ben~Q Baragiola, Saikat Guha, Guillaume Dauphinais, et~al.
\newblock Blueprint for a scalable photonic fault-tolerant quantum computer.
\newblock {\em Quantum}, 5:392, 2021.

\bibitem{PRXQuantum3030309}
Srikrishna Omkar, Seok-Hyung Lee, Yong~Siah Teo, Seung-Woo Lee, and Hyunseok Jeong.
\newblock All-photonic architecture for scalable quantum computing with {G}reenberger-{H}orne-{Z}eilinger states.
\newblock {\em PRX Quantum}, 3:030309, 2022.

\bibitem{fukui2023high}
Kosuke Fukui.
\newblock High-threshold fault-tolerant quantum computation with the gottesman-kitaev-preskill qubit under noise in an optical setup.
\newblock {\em Physical Review A}, 107(5):052414, 2023.

\bibitem{GKP_Analog}
Kosuke Fukui, Akihisa Tomita, and Atsushi Okamoto.
\newblock Analog quantum error correction with encoding a qubit into an oscillator.
\newblock {\em Phys. Rev. Lett.}, 119:180507, 2017.

\bibitem{GKP_photonloss}
Victor~V. Albert, Kyungjoo Noh, Kasper Duivenvoorden, Dylan~J. Young, R.~T. Brierley, Philip Reinhold, Christophe Vuillot, Linshu Li, Chao Shen, S.~M. Girvin, Barbara~M. Terhal, and Liang Jiang.
\newblock Performance and structure of single-mode bosonic codes.
\newblock {\em Phys. Rev. A}, 97:032346, 2018.

\bibitem{GKP_Gaussian}
Ben~Q. Baragiola, Giacomo Pantaleoni, Rafael~N. Alexander, Angela Karanjai, and Nicolas~C. Menicucci.
\newblock All-gaussian universality and fault tolerance with the gottesman-kitaev-preskill code.
\newblock {\em Phys. Rev. Lett.}, 123:200502, 2019.

\bibitem{GKP_imp}
Shunya Konno, Warit Asavanant, Fumiya Hanamura, Hironari Nagayoshi, Kosuke Fukui, Atsushi Sakaguchi, Ryuhoh Ide, Fumihiro China, Masahiro Yabuno, Shigehito Miki, Hirotaka Terai, Kan Takase, Mamoru Endo, Petr Marek, Radim Filip, Peter van Loock, and Akira Furusawa.
\newblock Logical states for fault-tolerant quantum computation with propagating light.
\newblock {\em Science}, 383(6680):289--293, 2024.

\bibitem{GKP_generate}
Daniel~J. Weigand and Barbara~M. Terhal.
\newblock Generating grid states from schr\"odinger-cat states without postselection.
\newblock {\em Phys. Rev. A}, 97:022341, 2018.

\bibitem{H-cat1}
H.~Jeong, M.~S. Kim, and Jinhyoung Lee.
\newblock Quantum-information processing for a coherent superposition state via a mixedentangled coherent channel.
\newblock {\em Phys. Rev. A}, 64:052308, 2001.

\bibitem{H-cat2}
Seung-Woo Lee and Hyunseok Jeong.
\newblock Near-deterministic quantum teleportation and resource-efficient quantum computation using linear optics and hybrid qubits.
\newblock {\em Phys. Rev. A}, 87:022326, 2013.

\bibitem{PhysRevLett125060501}
Srikrishna Omkar, Yong~Siah Teo, and Hyunseok Jeong.
\newblock Resource-efficient topological fault-tolerant quantum computation with hybrid entanglement of light.
\newblock {\em Phys. Rev. Lett.}, 125:060501, 2020.

\bibitem{H-cat_FTQC}
S.~Omkar, Y.~S. Teo, Seung-Woo Lee, and H.~Jeong.
\newblock Highly photon-loss-tolerant quantum computing using hybrid qubits.
\newblock {\em Phys. Rev. A}, 103:032602, 2021.

\bibitem{H-cat3}
Jaehak Lee, Nuri Kang, Seok-Hyung Lee, Hyunseok Jeong, Liang Jiang, and Seung-Woo Lee.
\newblock Fault-tolerant quantum computation by hybrid qubits with bosonic cat-code and single photons.
\newblock arXiv:2401.00450 [quant-ph], 2023.

\bibitem{fukui2024resource}
Kosuke Fukui and Peter van Loock.
\newblock Resource-efficient high-threshold fault-tolerant quantum computation with weak nonlinear optics.
\newblock {\em arXiv preprint arXiv:2412.16536}, 2024.

\bibitem{H-protocol1}
Kimin Park, Seung-Woo Lee, and Hyunseok Jeong.
\newblock Quantum teleportation between particlelike and fieldlike qubits using hybrid entanglement under decoherence effects.
\newblock {\em Phys. Rev. A}, 86:062301, 2012.

\bibitem{H-protocol2}
Hoi-Kwan Lau and Martin~B. Plenio.
\newblock Universal quantum computing with arbitrary continuous-variable encoding.
\newblock {\em Phys. Rev. Lett.}, 117:100501, 2016.

\bibitem{H-protocol3_QKD}
Youngrong Lim, Jaewoo Joo, Timothy~P. Spiller, and Hyunseok Jeong.
\newblock Loss-resilient photonic entanglement swapping using optical hybrid states.
\newblock {\em Phys. Rev. A}, 94:062337, 2016.

\bibitem{H-protocol4_repeater}
Marcel Bergmann and Peter van Loock.
\newblock Hybrid quantum repeater for qudits.
\newblock {\em Phys. Rev. A}, 99:032349, 2019.

\bibitem{H-protocol6}
Seongjeon Choi, Seok-Hyung Lee, and Hyunseok Jeong.
\newblock Teleportation of a multiphoton qubit using hybrid entanglement with a loss-tolerant carrier qubit.
\newblock {\em Phys. Rev. A}, 102:012424, 2020.

\bibitem{H-protocol5}
Soumyakanti Bose and Hyunseok Jeong.
\newblock Quantum teleportation of hybrid qubits and single-photon qubits using gaussian resources.
\newblock {\em Phys. Rev. A}, 105:032434, 2022.

\bibitem{H-protocol7}
Soumyakanti Bose, Jaskaran Singh, Ad\'an Cabello, and Hyunseok Jeong.
\newblock Long-distance entanglement sharing using hybrid states of discrete and continuous variables.
\newblock {\em Phys. Rev. Appl.}, 21:064013, 2024.

\bibitem{gane_H-cat1}
W~J Munro, K~Nemoto, and T~P Spiller.
\newblock Weak nonlinearities: a new route to optical quantum computation.
\newblock {\em New Journal of Physics}, 7(1):137, may 2005.

\bibitem{gane_H-cat2}
Hyunseok Jeong.
\newblock Using weak nonlinearity under decoherence for macroscopic entanglement generation and quantum computation.
\newblock {\em Phys. Rev. A}, 72:034305, 2005.

\bibitem{gene_Hybrid}
Hyukjoon Kwon and Hyunseok Jeong.
\newblock Generation of hybrid entanglement between a single-photon polarization qubit and a coherent state.
\newblock {\em Phys. Rev. A}, 91:012340, 2015.

\bibitem{PhysRevA76060301}
Jun-ichi Yoshikawa, Toshiki Hayashi, Takayuki Akiyama, Nobuyuki Takei, Alexander Huck, Ulrik~L. Andersen, and Akira Furusawa.
\newblock Demonstration of deterministic and high fidelity squeezing of quantum information.
\newblock {\em Phys. Rev. A}, 76:060301, 2007.

\bibitem{Kim1995Phase}
Myung~Shik Kim and Nobuyuki Imoto.
\newblock Phase-sensitive reservoir modeled by beam splitters.
\newblock {\em Phys. Rev. A}, 52:2401--2410, 1995.

\bibitem{exp_cat5}
Alexei Ourjoumtsev, Hyunseok Jeong, Rosa Tualle-Brouri, and Philippe Grangier.
\newblock Generation of optical ‘schr{\"o}dinger cats’ from photon number states.
\newblock {\em Nature}, 448(7155):784--786, 2007.

\bibitem{exp_cat3}
Hiroki Takahashi, Kentaro Wakui, Shigenari Suzuki, Masahiro Takeoka, Kazuhiro Hayasaka, Akira Furusawa, and Masahide Sasaki.
\newblock Generation of large-amplitude coherent-state superposition via ancilla-assisted photon subtraction.
\newblock {\em Phys. Rev. Lett.}, 101:233605, 2008.

\bibitem{exp_cat2}
Thomas Gerrits, Scott Glancy, Tracy~S. Clement, Brice Calkins, Adriana~E. Lita, Aaron~J. Miller, Alan~L. Migdall, Sae~Woo Nam, Richard~P. Mirin, and Emanuel Knill.
\newblock Generation of optical coherent-state superpositions by number-resolved photon subtraction from the squeezed vacuum.
\newblock {\em Phys. Rev. A}, 82:031802, 2010.

\bibitem{exp_cat1}
K.~Huang, H.~Le~Jeannic, J.~Ruaudel, V.~B. Verma, M.~D. Shaw, F.~Marsili, S.~W. Nam, E~Wu, H.~Zeng, Y.-C. Jeong, R.~Filip, O.~Morin, and J.~Laurat.
\newblock Optical synthesis of large-amplitude squeezed coherent-state superpositions with minimal resources.
\newblock {\em Phys. Rev. Lett.}, 115:023602, 2015.

\bibitem{exp_cat4}
Alexander~E Ulanov, Ilya~A Fedorov, Demid Sychev, Philippe Grangier, and AI~Lvovsky.
\newblock Loss-tolerant state engineering for quantum-enhanced metrology via the reverse hong--ou--mandel effect.
\newblock {\em Nature communications}, 7(1):11925, 2016.

\bibitem{exp_cat6}
Demid~V Sychev, Alexander~E Ulanov, Anastasia~A Pushkina, Matthew~W Richards, Ilya~A Fedorov, and Alexander~I Lvovsky.
\newblock Enlargement of optical schr{\"o}dinger's cat states.
\newblock {\em Nature Photonics}, 11(6):379--382, 2017.

\bibitem{surface}
Austin~G. Fowler, Matteo Mariantoni, John~M. Martinis, and Andrew~N. Cleland.
\newblock Surface codes: Towards practical large-scale quantum computation.
\newblock {\em Phys. Rev. A}, 86:032324, 2012.

\bibitem{XZZX}
J.~Pablo Bonilla~Ataides, David~K. Tuckett, Stephen~D. Bartlett, Steven~T. Flammia, and Benjamin~J. Brown.
\newblock The xzzx surface code.
\newblock {\em Nature Commun.}, 12(1), 2021.

\bibitem{XZZX_cat}
Andrew~S. Darmawan, Benjamin~J. Brown, Arne~L. Grimsmo, David~K. Tuckett, and Shruti Puri.
\newblock Practical quantum error correction with the xzzx code and kerr-cat qubits.
\newblock {\em PRX Quantum}, 2:030345, 2021.

\bibitem{surface_GKP}
Kyungjoo Noh, Christopher Chamberland, and Fernando~G.S.L. Brand\~ao.
\newblock Low-overhead fault-tolerant quantum error correction with the surface-gkp code.
\newblock {\em PRX Quantum}, 3:010315, 2022.

\bibitem{XZZX_GKP}
Jiaxuan Zhang, Yu-Chun Wu, and Guo-Ping Guo.
\newblock Concatenation of the gottesman-kitaev-preskill code with the xzzx surface code.
\newblock {\em Phys. Rev. A}, 107:062408, 2023.

\bibitem{PhysRevX10011058}
Arne~L. Grimsmo, Joshua Combes, and Ben~Q. Baragiola.
\newblock Quantum computing with rotation-symmetric bosonic codes.
\newblock {\em Phys. Rev. X}, 10:011058, 2020.

\bibitem{endo2022quantum}
Suguru Endo, Yasunari Suzuki, Kento Tsubouchi, Rui Asaoka, Kaoru Yamamoto, Yuichiro Matsuzaki, and Yuuki Tokunaga.
\newblock Quantum error mitigation for rotation symmetric bosonic codes with symmetry expansion.
\newblock {\em arXiv preprint arXiv:2211.06164}, 2022.

\bibitem{GKP_ion}
C.~Fl{\"u}hmann, T.~L. Nguyen, M.~Marinelli, V.~Negnevitsky, K.~Mehta, and J.~P. Home.
\newblock Encoding a qubit in a trapped-ion mechanical oscillator.
\newblock {\em Nature}, 566(7745):513--517, 2019.

\bibitem{GKP_sup}
P.~Campagne-Ibarcq, A.~Eickbusch, S.~Touzard, E.~Zalys-Geller, N.~E. Frattini, V.~V. Sivak, P.~Reinhold, S.~Puri, S.~Shankar, R.~J. Schoelkopf, L.~Frunzio, M.~Mirrahimi, and M.~H. Devoret.
\newblock Quantum error correction of a qubit encoded in grid states of an oscillator.
\newblock {\em Nature}, 584(7821):368--372, 2020.

\end{thebibliography}
\bibliographystyle{unsrt} 

\appendix*

\end{document}